\documentclass[preprint,showpacs,preprintnumbers,amssymb]{revtex4}

\usepackage{graphicx}
\usepackage{dcolumn}
\usepackage{bm}

\begin{document}

\title{Special relativity with a preferred frame and
the relativity principle: cosmological implications}
\author{Georgy I. Burde}
 \email{georg@bgu.ac.il}
 \affiliation{Alexandre Yersin Department of Solar Energy and Environmental Physics, Swiss Institute for Dryland Environmental and Energy Research, The Jacob Blaustein Institutes for Desert Research,
Ben-Gurion University of the Negev, Sede Boqer Campus, 84990,
Israel}
\begin{abstract}
The modern view, that there exists a preferred frame of reference related to the cosmic microwave background (CMB),
is in apparent contradiction with the principles of special relativity.
The purpose of the present study is to develop a counterpart of the special relativity theory that is consistent with  the existence of a preferred frame but, like the standard relativity theory, is based on the relativity principle and universality of the (\textit{two-way}) speed of light.
The synthesis of those seemingly incompatible concepts is possible at the expense of the freedom in assigning the \textit{one-way} speeds of light that exists in special relativity.
In the framework developed, a degree of anisotropy of the one-way velocity
acquires meaning of a characteristic of the really existing anisotropy caused by motion of an inertial frame relative to the preferred frame.
The anisotropic special relativity kinematics
is developed based on the first principles: (1) Space-time transformations between inertial frames leave the equation of anisotropic light propagation
invariant and (2) A set of the transformations possesses a group structure.
The Lie group theory apparatus is applied to define groups of space-time transformations between inertial frames. The correspondence principle, that the coordinate transformations should turn into the Galilean transformations in the limit of small velocities, and the argument, that the anisotropy parameter $k$ in a particular inertial frame is determined by its velocity relative to the preferred frame, are used to specify the transformations. The parameter of anisotropy $k$ becomes a variable which takes part in the transformations so that
the preferred frame
naturally arises as the frame with $k=0$.
The transformations between inertial frames obtained as the result of the analysis
do not leave the interval between two events invariant but modify it by a conformal factor.
Applying the consequences of the transformations to the problem of calculating
the CMB temperature distribution yields an equation in which the angular dependence coincides with that obtained on the basis of the standard relativity theory but the mean temperature is corrected by the terms second order in 
the observer velocity.
From conceptual point of view,
it eliminates the inconsistency of the usual approach when formulas of the standard special relativity
are applied to define effects caused by motion with respect to the preferred frame.
\end{abstract}



\maketitle

\section{Introduction}

Special relativity underpins nearly all of present day physics.
Lorentz invariance is one of the cornerstones of general relativity and other theories of fundamental physics.
However,
the discovery of the cosmic microwave background (CMB) radiation has shown that cosmologically a preferred system of reference does exist which is in apparent contradiction with
the principles of the special relativity theory.
Nevertheless, the formulas of special relativity are commonly used in cosmological context when there is a need to relate physical effects in the frames moving with respect to each other. Applying the Doppler effect and the light aberration equations based on the Lorentz transformations for calculating the CMB temperature anisotropies due to our galaxy's peculiar motion with respect to the CMB provides an example of such an approach.

The view, that there exists a preferred frame of reference, seems to unambiguously lead to the abolishment of the basic principles of the special relativity theory: the principle of relativity
and the principle of universality of the speed of light. The modern
versions of experimental tests of special relativity and the "test theories" of special relativity
\cite{Rob}--\cite{MS3}
presume that a preferred inertial reference frame ("rest" frame), identified with the CMB frame, is the only frame in which
the \textit{two-way} speed of light (the average speed from source to observer and back) is isotropic.
Furthermore, it seems that accepting the existence of a preferred frame forces one to abandon the group structure for the set of space-time transformations between
inertial frames. In the test theories, transformations between 'moving' frames are not considered, only a form of the transformation between a preferred frame and a particular moving frame is postulated.

The purpose of the present study is to develop a counterpart of the special relativity kinematics, that is consistent with the existence of a preferred frame but, like the standard relativity theory, is based on the universality of the (two-way) speed of light and the relativity principle. The group structure of a set of transformations between inertial frames is also preserved in the theory developed.
The reconciliation and synthesis of those concepts
with the  existence of a preferred frame
is possible at the expense of the freedom in assigning the \textit{one-way} speeds of light. The one-way speed of light is commonly considered as irreducibly conventional in view of the fact that it cannot be defined separately from the synchronization choice
(see, e.g., \cite{Ung1}-- \cite{Rizzi}).
Nevertheless, the analysis shows that, despite the inescapable entanglement between remote clock synchronization and one-way speed of light,
a specific value of the \textit{one-way speed of light and corresponding synchronization} are selected  from others in some objective way. In the framework developed, the argument that the anisotropy of the one-way speed of light in a particular inertial frame is due to its motion relative to the preferred frame, being combined  with the requirements of invariance of the equation of (anisotropic) light propagation and the group structure of a set of transformations between inertial frames, defines a specific value of the one-way speed of light for that frame. The parameter of anisotropy of the one-way speed of light $k$ becomes a variable that takes part in the group transformations. The preferred frame, commonly defined by that the propagation of light in that frame is isotropic,
is naturally present in the analysis as the frame with $k=0$ and it does not violate the relativity principle since the transformations from/to that frame are  not distinguished from other members of the group.

The space-time transformations between inertial frames derived as a result of the analysis differ from the Lorentz transformations. Since the theory is based  on the special relativity principles, it means that the Lorentz invariance is violated without violation of the relativistic invariance. The theory equations contain one undefined universal constant $q$ such that the case of $q=0$ corresponds to the standard special relativity with isotropic one-way speed of light in all inertial frames. The measurable effects following from the theory equations can provide estimates for $q$ and define deviations from the standard relativity that way.
Applying the theory to the problem of calculating
the CMB temperature distribution eliminates the inconsistency of the usual approach when formulas of the standard special relativity, which does not allow a preferred frame, are used to define effects caused by motion with respect to the preferred frame. The CMB temperature angular dependence predicted by the present theory coincides with that obtained on the basis of the standard relativity equations while the mean temperature is corrected by the terms second order in 
the observer velocity.

\subsection{Anisotropy of the one-way speed of light in special relativity}

The issue of anisotropy of the one-way speed of light is traditionally placed into the context of conventionality of distant simultaneity and clock synchronization \cite{Ung1}-- \cite{Rizzi}.
Simultaneity at distant space points
of an inertial system is defined by a clock synchronization that
makes use of light signals.
If a light ray is emitted from the master clock
and reflected off the remote clock one has a freedom to give the
reflection time $t$ at the remote clock any intermediate time in
the interval between the emission and reception times $t_0$ and
$t_R$ at the master clock
\begin{equation}\label{In1}
t=t_0+\epsilon(t_R-t_0);\quad 0<\epsilon <1
\end{equation}
where $\epsilon$ is Reichenbach's synchrony parameter
\cite{Reich}. The thesis that the value of the synchrony parameter
$\epsilon$ may be freely chosen in $0<\epsilon<1$ is known as the
conventionality of simultaneity. Reichenbach's "nonstandard"
synchronization reduces to the "standard" Einstein synchronization
when  $\epsilon=1/2$. Any choice of $\epsilon\neq 1/2$ corresponds
to assigning different one-way speeds of light signals in each
direction which must satisfy the condition that the average is
equal to $c$. Speed of light in each direction is
therefore\begin{equation}\label{In2} V_{\pm}=\frac{c}{1\pm
k_{\epsilon}},\quad k_{\epsilon}=2\epsilon -1
\end{equation}
If the described procedure is used for setting up throughout the
frame of a set of clocks using signals from some master clock
placed at the spatial origin, a difference in the standard and
nonstandard clock synchronization may be reduced to a change of
coordinates \cite{Ung1}-- \cite{Rizzi}
\begin{equation}\label{In3}
t=t^{(s)}+\frac{k_{\epsilon} x}{c},\quad x=x^{(s)}
\end{equation}
where $t^{(s)}=(t_0+t_R)/2$ is the time setting according to
Einstein (standard) synchronization procedure.

The analysis can be extended to the three dimensional case. If a
beam of light propagates (along straight lines) from a starting
point and through the reflection over suitable mirrors covers a
closed part the experimental fact is that the speed of light as
measured over closed part is always $c$ (\textit{Round-Trip Light
Principle}). In accordance with that experimental fact, if the speed of light is allowed to be
anisotropic it must depend on the direction of propagation  as \cite{And}, \cite{Min}
\begin{equation}\label{In4a}
V=\frac{c}{1+\mathbf{k_{\epsilon}}\mathbf{n}}=\frac{c}{1+k_{\epsilon}\cos
\theta_k}
\end{equation}
where $\mathbf{k_{\epsilon}}$ is a constant vector and $\theta_k$
is the angle between the direction of propagation $\mathbf{n}$ and
$\mathbf{k_{\epsilon}}$.
Similar to the
one-dimensional case, the law (\ref{In4a}) may be considered as a
result of the transformation from "standard" coordinatization of
the four-dimensional space-time manifold, with $k_{\epsilon}=0$,
to the "nonstandard" one with $k_{\epsilon}\neq 0$:
\begin{equation}\label{In5}
t=t^{(s)}+\frac{\mathbf{k_{\epsilon}}\mathbf{r}}{c},\quad
\mathbf{r}=\mathbf{r}^{(s)}
\end{equation}


There were several studies exploring the kinematics of the special
relativity theory when a definition of simultaneity other than
that used by Einstein is adopted. In particular, the
transformations, which are treated as replacing standard Lorentz
transformations of special relativity in the case of the
"nonstandard" synchronization (\ref{In1}) with $\epsilon\neq 1/2$,
have been repeatedly derived in the literature (see, e.g., \cite{Edw} -- \cite{Ung2}). Although somewhat
different assumptions (in addition to the common round-trip
light principle and the linearity assumption) are used in those studies, the
transformations derived, in fact, are identical -- they either coincide or become coinciding after a
 parameter change.
In what follows, those transformations will be called the
"$\epsilon$-Lorentz transformations", the name is due to
\cite{WinII}, \cite{Ung2}. The $\epsilon$-Lorentz transformations can be obtained from the standard Lorentz transformations by
a change of coordinates (\ref{In3}). Thus, the
$\epsilon$-Lorentz transformations are in fact the Lorentz
transformations of the standard special relativity represented
using the "nonstandard" coordinatization of the four-dimensional
space-time manifold. This might be expected in view of the fact
that the kinematic arguments used in the derivations of the
$\epsilon$-Lorentz transformations in the aforementioned works are
based on the assumption that, in the case of $\epsilon=1/2$, the
relations of the special relativity theory in its standard
formulation are valid.

It is commonly believed that, since the speed of light cannot be defined separately from the synchronization choice, the one-way speed of light is irreducibly conventional. Also, a possibility to introduce the
$\epsilon$-Lorentz transformations is considered as an illustration of conventionality of the one-way speed of light. Nevertheless, there are arguments showing that, despite the inescapable entanglement between remote clock synchronization and one-way
speed of light, a specific value of the \textit{one-way speed of light and the corresponding synchronization} can be distinguished from others.
In particular, it can be shown that the $\epsilon$-Lorentz transformations, usually considered as incorporating an anisotropy,
are in fact not applicable to the situation when there is an anisotropy in a physical system and that, in the  case of isotropy,  the particular case of the transformations corresponding to the isotropic one-way speed of light and Einstein synchronization (standard Lorentz transformations) is privileged.

The first point is that the $\epsilon$-Lorentz transformations do not satisfy the \textit{Correspondence Principle} unless the standard (Einstein) synchrony is used.
The correspondence principle
was taken by Niels Bohr as the guiding principle to discoveries in
the old quantum theory. Since then
the correspondence principle had germinated and
was considered as a guideline for the selection of new theories in physical science. In the context of special relativity, the correspondence principle
is traditionally mentioned  as a statement that Einstein's theory
of special relativity reduces to classical mechanics in the limit
of small velocities in comparison to the speed of light. Nevertheless, the correspondence principle
has not been properly used as a \textit{heuristic principle} in
developing the special relativity theory.

Being applied to the special relativity kinematics, the correspondence principle
implies that \textit{the transformations between inertial frames
should turn into the Galilean transformations in the limit of
small velocities}. Let us consider from this point of view the $\epsilon$-Lorentz transformations.
The $\epsilon$-Lorentz transformations between two arbitrary inertial reference frames $S$ and $S'$ in
the standard configuration, with the $y$- and $z$-axes of the two
frames being parallel while the relative motion with velocity $v$ is along the
common $x$-axis, being written in terms of $k_{\epsilon}$
(instead of $\epsilon$ as in \cite{Edw} -- \cite{Ung2}) take the forms
\begin{equation}
\label{Edw0} x=\frac{X-c T\beta
}{\sqrt{\left(1-k_{\epsilon}\beta\right)^2-\beta^2}},\quad c
t=\frac{c T\left(1-2k_{\epsilon}\beta\right)
-X\left(1-k_{\epsilon}^2\right)\beta
}{\sqrt{\left(1-k_{\epsilon}\beta \right)^2-\beta^2}}; \quad \beta=\frac{v}{c}
\end{equation}
Here the space and time coordinates in $S$ and $S'$ are denoted respectively as $\{X,Y,Z,T\}$ and $\{x,y,z,t\}$
and it is implied that clocks in the frames
$S$ and $S'$ are synchronized according to the anisotropy degree
$k_{\epsilon}$.
In the limit
of $\beta\rightarrow 0$, the formula for transformation of the coordinate $x$ (first equation of (\ref{Edw0})) turns into
\begin{equation}\label{Edw01}
x=X-\beta c T+k_{\epsilon} \beta X
\end{equation}
which is not coinciding with
the formula for transformation of the coordinate $x$
of the Galilean transformation
\begin{equation}\label{Edw02}
x=X- v T=X-\beta c T
\end{equation}

It should be noted that the relations $t=T$, $y=Y$ and $z=Z$,
which are commonly included into the system of equations called
the Galilean transformations, are not required to be valid in the
limit of small velocities. The fact that the first order (in $v$)
terms do not appear in those relations does not obligatory imply
that they should be absent in the first order approximations of
the special relativity formulas. In particular, if an expansion of
the Lorentz transformations with respect to $\beta=v/c$ is made
the first order term arises in the expansion of the time
transformation (see, for example, discussion in \cite{Ghosal},
\cite{Baierlein}). So only the relation (\ref{Edw02}), which
\textit{does} contain the first order term, provides a reliable
basis for applying the correspondence principle.

The additional,
as compared  with (\ref{Edw02}),
term appearing in the small velocity limit (\ref{Edw01}) of the $\epsilon$-Lorentz transformations includes the synchronization parameter and light speed which are alien to
the framework of the Galilean kinematics. Thus, the "$\epsilon$-Lorentz transformations" (\ref{Edw0}) do not satisfy the correspondence principle unless $k_{\epsilon}=0$ which means that applying the correspondence principle singles out the isotropic one-way speed of light and Einstein synchrony.

The next point is that the "$\epsilon$-Lorentz transformations"
are applicable only to the situation when there is no anisotropy in a physical system, for the reason that they leave the interval between two events invariant.
Invariance of the interval
is commonly considered as an integral part of the physics of special relativity which is
used as a starting point for derivation of the coordinate transformations between inertial frames. Nevertheless, invariance of the interval
is not a straightforward consequence of the basic principles of the theory. The two principles constituting the conceptual basis of the special relativity,  the \textit{principle of relativity} which states the equivalence of all inertial frames as regards the formulation of the laws of physics and \textit{universality of the speed of light} in inertial frames,
taken together
lead to the condition of \textit{invariance of the equation of light propagation} with respect to the coordinate transformations between inertial frames.
Thus, in general, not the invariance of the interval but invariance of the equation of light propagation should be a starting point for derivation of the transformations between inertial frames.
Therefore the use of the interval invariance is usually preceded
by a proof of its validity (see, e.g., \cite{Pauli}, \cite{LL}) based on
invariance of the equation of light propagation. However, those proofs are not valid if an anisotropy is present.

In such proofs, two reference frames $S$ and $S'$ in a standard configuration, with $S'$ moving with respect to $S$ with velocity $v$, are considered. First, it is stated that, under the assumption of linearity of coordinate transformations between the frames, the two equations $ds^2=0$ and $dS^2=0$, with $ds^2$ and $dS^2$ being the intervals between two events in the frames $S'$ and $S$,
can be valid only if $ds^2=\lambda (v)dS^2$ where $\lambda (v)$ is an
arbitrary function.
Next, the third frame $S''$ moving with velocity $(-v)$ with
respect to $S'$ (being at rest to $S$) is introduced and the transformations are applied once more
\begin{equation}
(ds'')^2=dS^2=\lambda (-v)ds^2=\lambda(-v)\lambda(v)dS^2\nonumber
\end{equation}
which results in
\begin{equation}
\label{Argi}\lambda(-v)\lambda(v)=1
\end{equation}
Applying the same arguments to a transversal coordinate yields
\begin{eqnarray}
&&y'=\kappa(v)y,\; y''=y=\kappa (-v)y'=\kappa (-v)\kappa(v)y\nonumber\\
&&\label{Int1} \kappa (-v)\kappa(v)=1
\end{eqnarray}
where $\kappa(v)$ corresponds to the change of the transverse dimensions of the rod. It is concluded that, \textit{for reasons of symmetry},  it should be independent of the direction of the velocity
\begin{equation}
\kappa(-v)=\kappa(v)\nonumber
\end{equation}
which, together with (\ref{Int1}), leads to
\begin{equation}
\kappa(v)=1,\quad \lambda (v)=\kappa(v)^2=1\nonumber
\end{equation}

However, \textit{the symmetry arguments are not valid} if an
anisotropy in the physical system is present. As a physical
phenomenon it influences all the processes so that any effects due
to movement of frame $S'$ relative to $S$ in some direction
are not equivalent to those due to
movement of frame $S''$ relative to $S$ in the opposite direction.
Therefore
$\kappa(v)\neq \kappa(-v)$ -- only the relation
$\kappa(v)\kappa(-v)=1$ or $\kappa(v)=1/\kappa(-v)$ should
be valid. Thus, in the presence of
the anisotropy, the interval should not be invariant. Moreover, $\lambda (v)\neq 1$ implies that strict invariance should be replaced by \textit{conformal invariance}.

The "$\epsilon$-Lorentz transformations" leave the interval between two events invariant (the
derivations of the $\epsilon$-Lorentz transformations in the
aforementioned works \cite{Edw} -- \cite{Ung2} are also based on invariance of the interval or equivalent assumptions)
and therefore they are applicable only to the case of no anisotropy.
Although, even in that case, the anisotropic one-way speed of light satisfying equation (\ref{In4a}) with $k_{\epsilon}\neq 0$ and the corresponding $\epsilon$-Lorentz transformations are mathematically acceptable, it is conceptually inconsistent to apply the transformations with anisotropic speed of light to isotropic situation. Moreover, the transformations themselves are physically inconsistent since they do not satisfy the correspondence principle.
Thus, the value of  $k_{\epsilon}=0$ is privileged in some objective way if no anisotropy is present in a physical system.

It follows from the above discussion that, in the case of an \textit{anisotropic} system, there should also exist a privileged value of the light-speed anisotropy parameter $k_{\epsilon}$ selected by the size of the anisotropy.

\subsection{Conceptual framework}

The special relativity kinematics applicable to an anisotropic system
should be developed based on the first principles of special relativity but without refereeing to the relations of the standard relativity theory.
The principles constituting the conceptual basis of special relativity, the relativity principle, according to which physical laws should have the same forms in all inertial frames, and the universality of the speed of light in inertial frames,
lead to the requirement of
invariance of the equation of light propagation with respect to the coordinate transformations between inertial frames. In the present context, it should be invariance of the equation of propagation of light which incorporates the anisotropy of the one-way speed of light, with the law of variation of the speed with direction consistent with the experimentally verified round-trip light principle, as follows
\begin{equation}\label{In4}
V=\frac{c}{1+\mathbf{k}\mathbf{n}}=\frac{c}{1+k\cos \theta_k}
\end{equation}
where $\mathbf{k}$ is a (constant) vector characteristic of the anisotropy.
The change of notation, as compared with (\ref{In4a}), from
$k_{\epsilon}$ to $k$ is intended to indicate that $\mathbf{k}$ is
a parameter value corresponding to the size of the really existing anisotropy
while $k_{\epsilon}$ defines the anisotropy in the one-way speeds
of light due to the nonstandard synchrony equivalent to the coordinate change
(\ref{In5}).
The anisotropic
equation of light propagation incorporating the law (\ref{In4}) has the form (see Appendix A)
\begin{equation}\label{Sec3.1_Int}
ds^2=c^2 dt^2-2k c\; dtdx-(1-k^2)dx^2-dy^2-dz^2=0
\end{equation}
where $(x,y,z)$ are coordinates and $t$ is time. It is assumed that the $x$-axis is chosen to be along the anisotropy vector $\mathbf{k}$. Note that although the form (\ref{Sec3.1_Int}) is usually attributed to the
one-dimensional formulation it can be shown that,
in the three-dimensional case, the equation has the same form if
the anisotropy vector
$\mathbf{k}$ is directed along the $x$-axis (see Appendix A).





Further, in the development of the anisotropic relativistic
kinematics, a number of other physical requirements,
associativity, reciprocity and so on are to be satisfied which all
are covered by the condition that the transformations between
the frames form a group.
Thus, the group property should be taken as another first
principle.
The formulation based on the invariance and group property suggests using the \textit{Lie group theory} apparatus for defining groups of space-time transformations between inertial frames.

At this point, it should be clarified that there can exist two different cases: (1)
The size of anisotropy does not depend on the observer motion and so is the same in all inertial frames; (2) The anisotropy is due to the observer motion with respect to a preferred frame and so
the size of anisotropy varies from frame to frame.

In the first case, the group of transformations leaving equation (\ref{Sec3.1_Int}) invariant should be defined under the condition that $k$ is a constant. If two arbitrary inertial reference frames $S$ and $S'$ in
the standard configuration , with the space and time coordinates in $S$ and $S'$
denoted respectively as $\{X,Y,Z,T\}$ and $\{x,y,z,t\}$, are considered, then the equation of anisotropic light propagation in the frames $S$ and $S'$ takes the forms
\begin{eqnarray}\label{Int21}
&&c^2 dT^2-2k c\; dTdX-(1-k^2)dX^2-dY^2-dZ^2=0,\\
&&\label{Int21a}c^2 dt^2-2k c\;
dtdx-\left(1-k^2\right)dx^2-dy^2-dz^2=0
\end{eqnarray}
The transformations from $\{X,Y,Z,T\}$ to $\{x,y,z,t\}$ are sought such that equation (\ref{Int21}) is converted into (\ref{Int21a}) under the transformations. Groups of transformations defined within this framework are considered in \cite{burde}.

The second case is relevant to the purpose of the present study -- developing the special relativity kinematics consistent with the existence of a preferred frame. Since the anisotropy parameter varies from frame to frame, the equations of light propagation in the frames $S$ and $S'$ are
\begin{eqnarray}\label{Int22}
&&c^2 dT^2-2K c\; dTdX-(1-K^2)dX^2-dY^2-dZ^2=0,\\
&&\label{Int22a}c^2 dt^2-2k c\;
dtdx-\left(1-k^2\right)dx^2-dy^2-dz^2=0
\end{eqnarray}
where $k$ differs from $K$. Thus, the anisotropy parameter becomes a variable which takes part in the transformations and so groups of transformations in \textit{five} variables $\{x,y,z,t,k\}$ which convert  (\ref{Int22}) into (\ref{Int22a}) are sought.
In such a framework, the preferred frame,
commonly defined by that the propagation of light in that frame is isotropic,
naturally arises as the frame in which $k=0$. However, it does not violate the relativity principle since the transformations from/to that frame are  not distinguished from other members of the group. Nevertheless, the fact, that the anisotropy of the one-way speed of light in an arbitrary inertial frame is due to motion of that frame relative to the preferred frame,
is a part of the paradigm  which is used in the analysis,

The procedure of obtaining the transformations using the Lie group infinitesimal technique consists of the following steps:  (1) The infinitesimal invariance condition is applied to the equation of light propagation which  yields determining equations for the infinitesimal group generators; (2) Having the group generators defined the finite transformations are determined as solutions of the Lie equations; (3) The group
parameter is related to physical parameters using some obvious
conditions. The transformations between inertial frames derived in
such a way, both for a constant anisotropy parameter and a variable anisotropy parameter, contain a scale factor
and thus do not leave the interval
between two events invariant but modify it by a conformal factor
(square of the scale factor).

Conformal invariance can be expected if the transformations are
derived using the invariance of the equation of light propagation
and group property, like as the conformal group arises if invariance of the electrodynamic equations is studied \cite{Cunn}, \cite{Bate}  (see reviews
\cite{Fult}, \cite{Kast} for further developments).
Transformations which conformally modify
Minkowski metric have been introduced in the context of the special relativity kinematics in the presence of space anisotropy in \cite{Bogoslov1}, \cite{Bogoslov2} (two papers from the series) and \cite{Sonego} (see also \cite{Lalan}). As distinct from the present framework,
in the works \cite{Bogoslov1} -- \cite{Sonego},
the assumption that the form of the metric changes by a "conformal" factor is \textit{imposed}.
Therefore any values of the conformal factor are permissible and in particular, it may be equal to one which
reduces the transformations to the Lorentz transformations or to
the $\epsilon$-Lorentz transformations if a nonstandard clock synchronization is accepted  (see \cite{burde} for a more detailed discussion of the works \cite{Bogoslov1} -- \cite{Sonego}).

Although, in the present analysis, the
conformal invariance of the metric is not imposed but \textit{arises} as an intrinsic feature of special relativity based on invariance of the
anisotropic equation of light propagation and the group property, the conformal factor includes an arbitrary element. Therefore, in order to
complete the construction of the anisotropic relativistic
kinematics, one needs to relate the conformal factor to the parameter $k$ (or $K$ in the case when this parameter varies from frame to frame)
characterizing the anisotropy of the light speed.
The \textit{correspondence principle},
according to which the transformations between inertial frames
should turn into the Galilean transformations in the limit of
small velocities, is used for this purpose. (As a matter of fact, the isotropic case provides a guiding example for this: it is the correspondence principle that assigns a preferred status to $k_{\epsilon}=0$ if the scale factor $\lambda=1$.)
It is evident that validity of this principle should not depend on whether propagation of light is assumed to be isotropic or anisotropic since, in the framework of the Galilean kinematics, there is no place for the issues of light speed and its anisotropy.
Thus, the anisotropic
special relativity kinematics is developed using the following
three principles: \textit{the transformations form a group, leave the equation of anisotropic light propagation invariant and satisfy the correspondence principle}. (Applying these principles to the case of isotropic light propagation yields the standard Lorentz transformations of special relativity, see Appendix B).

Applying those principles for deriving transformations between inertial frames, in the case when the anisotropy parameter $k$ does not vary from frame to frame \cite{burde},
yields the transformations which include
the scale (conformal) factor $\lambda$
in such a way that the value $\lambda=1$ is
not allowed unless $k=0$.
Therefore, as distinct from the
"$\epsilon$-Lorentz transformations", the transformations defined within that framework
cannot be converted into the standard Lorentz transformations by a
synchrony change.
Measurable effects that arise as the consequences of the
transformations do not depend on the synchronization choice and so
allow, in principle, to determine the size of the anisotropy. It means that  a privileged value of the one-way speed of light and the corresponding synchronization can be objectively selected by a size of the anisotropy.

In the case of
the variable anisotropy parameter $k$, which is the subject of the present study, derivation of transformations between inertial frames, although being based on the same first principles as in the case of a constant $k$, differs conceptually and methodologically from that case.
First, the fact, that now groups of transformations in five variables $\{x,y,z,t,k\}$ are sought, changes both the derivation procedure and the resulting transformations. Further, since the law of variation of the anisotropy parameter from frame to frame is not completely defined by the determining equations, the scale (conformal) factor contained in the transformations includes an undefined function of the group parameter. The conceptual argument, that the size of anisotropy of the one-way speed of light in an arbitrary inertial frame depends on its velocity relative to the preferred frame,
allows to specify the transformations. As the result, the specified transformations include, instead of an arbitrary function, only one undefined universal parameter.

The only preferred frame one may think of is of course the cosmological frame in which the microwave background radiation is isotropic.
Applying the consequences of the anisotropic relativity transformations developed in the present analysis to the problem of calculating
the CMB temperature distribution (as observed in the frame moving relative to CMB) yields the formula which differs from that obtained using equations of the standard relativity theory. The angular dependence appears to be the same but the mean temperature is corrected and  the corrections are of the order of $(\bar{v}/c)^2$ where $\bar{v}$ is the observer velocity with respect to the CMB.
The formula for the Doppler frequency shift of the present theory can be also applied to the case when an observer in a frame moving with respect to the CMB (Earth) receives light from an object (galaxy) which is also moving with respect to that preferred frame.

The paper is organized, as follows. In Section 2, the method is outlined and the coordinate transformations between inertial frames incorporating anisotropy of the light propagation, with the anisotropy parameter varying from frame to frame, are derived. In Section 3, the transformations are specified using the argument that the anisotropy of the light propagation is due to the observer motion with respect to the preferred frame. Consequences of the transformations and corresponding measurable effects are considered in Section 4. Cosmological implications are discussed in Section 5. Some concluding remarks are furnished in Section 6. In Appendix A, a derivation of the equation of light
propagation incorporating the anisotropy of light speed in the
general three-dimensional form (\ref{In4}) is presented. In Appendix B, the method is applied to the case of no anisotropy and it is shown that
it yields the standard Lorentz transformations.

\section{Transformations between inertial frames in the presence of anisotropy}

In this section, groups of transformations between inertial frames that leave the equation for
light propagation, incorporating the anisotropic law (\ref{In4}), form-invariant are defined. The parameter of anisotropy $k$ is allowed to vary from frame to
frame which, in particular, implies that there exists a preferred
frame in which the speed of light is isotropic.

Consider two arbitrary inertial reference frames $S$ and $S'$ in
the standard configuration with the $y$- and $z$-axes of the two
frames being parallel while the relative motion is along the
common $x$-axis. The space and time coordinates in $S$ and $S'$
are denoted respectively as $\{X,Y,Z,T\}$ and $\{x,y,z,t\}$. The
velocity of the $S'$ frame along the positive $x$ direction in
$S$, is denoted by $v$. It is assumed that the frame $S'$ moves relative to $S$ along
the direction determined by the vector $\mathbf{k}$ from
(\ref{In4}). This assumption is justified by that one
of the frames in a set of frames with different values of $k$
is a preferred frame, in which $k=0$, so that the transformations must include, as a particular case, the transformation to that preferred frame. Since the anisotropy is attributed to the fact of motion with respect to the
preferred frame it is expected that the axis of
anisotropy is along the direction of motion (however, the direction of the anisotropy vector  can be both coinciding and opposite to that of velocity).

Transformations between the frames are
derived based on the following first principles: \textit{invariance of the equation of light
propagation} (underlined by the relativity principle), \textit{group property} and the \textit{correspondence principle}. Note that the group property is used not as in
the traditional analysis which commonly proceeds along the lines initiated by \cite{Ignatowski}
and \cite{FrRo} which are based on the linearity assumption and
relativity arguments. The difference can be seen from the derivation of the standard Lorentz transformations in Appendix B.

\medskip

\noindent \textit{Invariance of the equation of light propagation}.
The equations for
light propagation in the frames $S$ and $S'$ are
\begin{eqnarray}\label{LPPref}
&& c^2 dT^2-2K c\; dTdX-(1-K^2)dX^2-dY^2-dZ^2=0,\\
&&\label{LPPref1}c^2 dt^2-2k c\;
dtdx-\left(1-k^2\right)dx^2-dy^2-dz^2=0
\end{eqnarray}
where the anisotropy parameters $K$ and $k$ in the frames $S$ and $S'$ are different. The relativity principle implies that the transformations of variables from $\{X,Y,Z,T,K\}$ to $\{x,y,z,t,k\}$ leave the form of the equation of light propagation invariant so that (\ref{LPPref}) is converted into (\ref{LPPref1}) under the transformations.

\medskip

\noindent\textit{Group property}.
The transformations between inertial frames form a one-parameter group
with the group parameter $a=a(v)$ (such that $v\ll 1$ corresponds
to $a\ll 1$):
\begin{eqnarray}\label{2.1}
&& x=f(X,Y,Z,T,K;a), \quad y=g(X,Y,Z,T,K;a), \quad
z=h(X,Y,Z,T,K;a),\nonumber \\
&& t=q(X,Y,Z,T,K;a); \quad k=p(K;a)
\end{eqnarray}
Remark that $k$ is a transformed variable taking part in the group transformations.
Based on the symmetry arguments it is assumed that the
transformations of the variables $x$ and $t$ do not involve the
variables $y$ and $z$ and vice versa:
\begin{equation}\label{2.4K}
x=f(X,T,K;a), \; t=q(X,T,K;a),\quad y=g(Y,Z,K;a), \; z=h(Y,Z,K;a);\quad k=p(K;a)
\end{equation}

\medskip

\noindent \textit{Correspondence principle}.
The correspondence principle requires that, in the limit
of small velocities $v\ll c$ (small values of the group parameter
$a\ll 1$), the formula for transformation of the coordinate $x$
turns into that of the Galilean transformation:
\begin{equation}\label{GalK}
x=X- v T
\end{equation}
Remark that the small $v$ limit is not influenced by the presence of anisotropy of the light propagation. It is evident that there should be no traces of light anisotropy in that limit, the issues of the light speed and its anisotropy are alien to the framework of Galilean kinematics.


The \textit{group property} and the requirement of \textit{invariance of the
equation of light propagation}  suggest applying the infinitesimal Lie technique
 (see, e.g., \cite{Bluman}, \cite{Olver}).
The infinitesimal transformations corresponding to (\ref{2.4K}) are introduced, as follows
\begin{eqnarray}\label{2.5K}
&&x\approx X+\xi(X,T,K)a, \quad t\approx T+\tau(X,T,K)a, \nonumber \\
&&y\approx Y+\eta(Y,Z,K)a, \quad z\approx Z+\zeta(Y,Z,K)a, \quad k \approx K+a \chi (K)
\end{eqnarray}
and equations (\ref{LPPref}) and (\ref{LPPref1}) are used to derive determining equations for
the group generators $\tau(X,T,K)$, $\xi(X,T,K)$, $\eta(Y,Z,K)$,
$\zeta(Y,Z,K)$ and $\chi (K)$. The finite group transformations are found then by solving the Lie equations with proper initial conditions

The \textit{correspondence principle} can be applied to specify partially the infinitesimal group generators.
Equation (\ref{GalK}) is used to calculate the group generator
$\xi(X,T)$, as follows
\begin{equation}\label{2.6}
\xi=\left(\frac{\partial x}{\partial a}\right)_{a=0}=
\left(\frac{\partial \left(X-v(a) T\right)}{\partial
a}\right)_{a=0}=-b T; \qquad b=v'(0)
\end{equation}
It can be set $b=1$ without loss of generality since
this constant can be eliminated by redefining the group parameter. Thus, the generator $\xi$ is defined by
\begin{equation}\label{xiAnizK} \xi=-T
\end{equation}
Then substituting the infinitesimal transformations
(\ref{2.5K}), with $\xi$ defined by (\ref{xiAnizK}), into equation  (\ref{LPPref1}) with subsequent linearizing with respect to $a$ and using equation (\ref{LPPref}) to eliminate $dT^2$ yields
\begin{eqnarray}\label{DEAnisK}
&&\left(-K c^2\tau_X
+\left(1-K^2\right)\left(K+c\tau_T\right)+\chi\left(K\right) c
K\right)dX^2  \nonumber \\
&&+c \left(c^2\tau_X+c K \tau_T+1+K^2-\chi\left(K\right) c\right)dX
dT  \nonumber \\
&&+\left(K+c\tau_T-c\eta_Y\right)dY^2
+\left(K+c\tau_T-c\zeta_Z\right)dZ^2
-c\left(\eta_Z+\zeta_Y\right)dY dZ=0
\end{eqnarray}
where subscripts denote differentiation with respect to the
corresponding variable. In view of arbitrariness of the
differentials $dX$, $dY$, $dZ$ and, $dT$, the equality (\ref{DEAnisK})
can be valid only if the coefficients of all the monomials in
(\ref{DEAnisK}) vanish which results in an overdetermined system of
determining equations for the group generators.

The generators $\tau$, $\eta$ and $\zeta$ found from the
determining equations yielded by (\ref{DEAnisK}) are
\begin{equation}\label{GenAnisK}
\tau=-\frac{1-K^2-\chi\left(K\right) c}{c^2}X-\frac{2K}{c}T+c_2;,\quad
\eta=-\frac{K}{c}Y+\omega Z+c_3,\;\zeta=-\frac{K}{c}Z-\omega Y+c_4
\end{equation}
where $c_2$, $c_3$ and $c_4$ are arbitrary constants. The
common kinematic restrictions that one event is the spacetime
origin of both frames and that the $x$ and $X$ axes slide along
another can be imposed to make the constants $c_2$, $c_3$ and
$c_4$ vanishing (space and time shifts are eliminated). In
addition, it is required that the $(x,z)$ and $(X,Z)$ planes
coincide at all times which results in $\omega=0$ and so excludes
rotations in the plane $(y,z)$.

The finite transformations are determined by solving the Lie equations which, after rescaling the group parameter as $\hat a=a/c$ together with
$\hat \chi=\chi c$ and omitting hats afterwards, take the forms
\begin{eqnarray}
\label{LEAnis0} &&\frac{d k(a)}{d
a}=\chi\left(k\left(a\right)\right);\quad  k(0)=K,\\
\label{LEAnis1K}&&\frac{d x(a)}{d a}=-c t(a), \quad \frac{d \left(c t\left(a\right)\right)}{d
a}=-\left(1-k\left(a\right)^2-\chi\left(k\left(a\right)\right)\right)x(a)-2 k(a) c t\left(a\right),\\
\label{LEAnis2P}&& \frac{d y(a)}{d a}=-k(a) y(a),\quad \frac{d
z(a)}{d a}=-k(a) z(a);\\
\label{ICK}&&x(0)=X,\; t(0)=T,\; y(0)=Y,\; z(0)=Z.
\end{eqnarray}
Because of the arbitrariness of $\chi\left(k\left(a\right)\right)$, the solution of the system of equations
(\ref{LEAnis0}), (\ref{LEAnis1K}) and (\ref{LEAnis2P}) contains an arbitrary function $k(a)$. Using (\ref{LEAnis0})
to replace $\chi\left(k\left(a\right)\right)$ in the second
equation of (\ref{LEAnis1K}) we obtain solutions of equations
(\ref{LEAnis1K}) subject to the initial conditions (\ref{ICK}) in the form
\begin{eqnarray}\label{txAnisP}
&& x=R\left(X \left(\cosh{a}+K\sinh{a}\right)-cT
\sinh{a}\right),\\
&& \label{txAnis1P} c t=R\Bigl(c T\left(
\cosh{a}-k\left(a\right)\sinh{a}\right)\nonumber\\
&&-X\left(\left(1-K k\left(a\right)\right)
\sinh{a}+\left(K-k\left(a\right)\right)\cosh{a}\right)\Bigr)
\end{eqnarray}
where $R$ is defined by
\begin{equation}\label{VarPhi}
R=e^{-\int_0^a{k(\alpha)d\alpha}}
\end{equation}
To complete the derivation of the transformations the group
parameter $a$ is to be related to the velocity $v$ using the
condition
\begin{equation}\label{2.8aK}
x=0 \quad \mathrm{for} \quad X=v T
\end{equation}
which yields
\begin{equation}\label{GrPAnisP}
a=\frac{1}{2}\ln{\frac{1+\beta - K\beta}{1-\beta-K\beta}};\qquad
\beta=\frac{v}{c}
\end{equation}
Substituting (\ref{GrPAnisP}) into (\ref{txAnisP}) and
(\ref{txAnis1P}) yields
\begin{eqnarray}\label{TrAnisP}
&&
x=\frac{R}{\sqrt{\left(1-K\beta\right)^2-\beta^2}}\left(X-c
T \beta\right),\nonumber \\
&&ct=\frac{R}{\sqrt{\left(1-K\beta\right)^2-\beta^2}}\left(c
T
\left(1-K\beta-k\beta\right)-X\left(\left(1-K^2\right)\beta+K-k\right)\right)
\end{eqnarray}
where $k$ is the value of $k(a)$ calculated for $a$ given by (\ref{GrPAnisP}).

 Solving equations
(\ref{LEAnis2P}) and using (\ref{GrPAnisP}) in the result yields
\begin{equation}\label{YZAnis2P}
y=R Y,\quad z=R Z
\end{equation}

Calculating the interval
\begin{equation}\label{IntAnisP}
ds^2=c^2 dt^2-2k c\; dtdx-(1-k^2)dx^2-dy^2-dz^2
\end{equation}
with (\ref{TrAnisP}) and (\ref{YZAnis2P}) yields
\begin{equation}\label{TrIntAnisP}
ds^2=R^2 dS^2,\quad dS^2=c^2 dT^2-2K c\;
dTdX-(1-K^2)dX^2-dY^2-dZ^2
\end{equation}
Thus, in the case when the anisotropy exists, the interval
invariance is replaced by \textit{conformal invariance} with the
conformal factor dependent on the relative velocity of the frames
and the anisotropy degree.

Although the group properties of the transformation defined by
equations (\ref{VarPhi}) -- (\ref{YZAnis2P}) are guaranteed by the
fact that they have been derived as solutions of the Lie equations
(\ref{LEAnis0})-- (\ref{LEAnis2P}) it is instructive to make a
straightforward check of the fact that the transformations obey
the basic group properties. To do this one may introduce, in
addition to the frames $S$ and $S'$, the third inertial frame
$S''$ with the space-time variables $(x_1,y_1,z_1,t_1)$ which
moves relative to $S'$ with the velocity $v_1$. The
transformations from $S'$ to $S''$ are given by the same equations
(\ref{TrAnisP}) and (\ref{YZAnis2P})
 but with $(x_1,y_1,z_1,t_1)$ replacing
$(x,y,z,t)$, $(x,y,z,t)$ replacing $(X,Y,Z,T)$, $\beta_1=v_1/c$
replacing $\beta$, $a_1$ replacing $a$, $k$ replacing $K$ and
$k_1=k(a_1)$ replacing $k$. Here $a_1$ is given by equation
(\ref{GrPAnisP}) with $K$ replaced by $k$ and $\beta$ replaced by
$\beta_1$. It is readily checked that substituting the
transformation formulas for $(x,y,z,t)$ into the transformation
formulas for $(x_1,y_1,z_1,t_1)$ yields again the formulas
(\ref{TrAnisP}) and (\ref{YZAnis2P}) but with $a$ and $\beta$ replaced by
$a_2$ and $\beta_2$, where $\beta_2=v_2/c$ is the velocity of $S''$ with respect to $S$ and $a_2=a_1+a$ is the corresponding value of the group parameter. The three velocities $v_1$, $v_2$ and $v$ are related by
\begin{equation}\label{GrProp}
\beta_1=\frac{\beta_2-\beta}{1- K
\beta+\beta_2(K^2\beta-K-\beta)+k(\beta_2-\beta)}
\end{equation}
Equation (\ref{GrProp}) could alternatively be obtained
from the relation $a_2=a_1+a$, in accordance with the basic group
property,  using a properly specified equation (\ref{GrPAnisP}).

Considering inverse transformations from the frame $S'$ to $S$ one
has to take into account that, in the presence of the light speed
anisotropy, the reciprocity principle is modified \cite{WinII},
\cite{Ung1}. The reasoning behind this is that all speeds are to
be affected by the anisotropy of the light speed since the speeds
are timed by their coincidences at master and remote clocks, and
the latter are altered. Therefore the relative velocity $v_{-}$ of
$S$ to $S'$ is not equal to the relative velocity $v$ of $S'$ to
$S$. The modified reciprocity relation can be derived with the use
of equation (\ref{In5}) which allows to relate the velocity
measured by the clocks synchronized with anisotropic light speed
to the velocity $v_s$ measured by the clocks synchronized with
isotropic light speed -- the latter is not dependent on the
direction. Using equation (\ref{In3}), which for
$\mathbf{k_{\epsilon}}$ directed along the $x$-axis is equivalent to (\ref{In5}),
one obtains
\begin{equation}\label{VelRec1Ka}
\frac{dx}{dt}=\frac{dx}{dt_s+\frac{k_{\epsilon}
dx}{c}}=\frac{\frac{dx}{dt_s}}{1+\frac{k_{\epsilon}}{c}\frac{dx}{dt_s}}
\end{equation}
If this equation is used to calculate the velocity $v$ of the frame $S'$ as measured by an observer in the frame $S$, then $k_{\epsilon}$, $dx/dt$ and $dx/dt_s$ are to be replaced by $K$, $dX/dT=v$ and $dX/dT_s=v_s$ respectively while, for calculating the velocity $v_{-}$ of the frame $S$ as measured in $S'$, the quantities $k_{\epsilon}$, $dx/dt$ and $dx/dt_s$ are to be replaced by $k$, $-v_{-}$ and $-v_s$, as follows
\begin{equation}\label{VelRec1Kb}
v=\frac{v_s}{1+\frac{K}{c}v_s}, \quad
(-v_{-})=\frac{(-v_s)}{1+\frac{k}{c}(-v_s)}
\end{equation}
 Eliminating the velocity $v_s$ from equations (\ref{VelRec1Kb}) yields
\begin{equation}\label{VelRec1Kc}
v_{-}=\frac{c v}{c-(k+K) v}\quad \Rightarrow \quad \beta_{-}=\frac{v_{-}}{c}=\frac{\beta}{1-(k+K) \beta}
\end{equation}
So, according to the modified reciprocity principle, the group
parameter value corresponding to the inverse transformation is
calculated from (\ref{GrPAnisP}) but with $K$ replaced by $k$ and $\beta$ replaced by
($-\beta_{-}$), as follows
\begin{equation}\label{GrPAnisI}
a_{-}=\frac{1}{2}\ln{\frac{1-\beta_{-} +
k\beta_{-}}{1+\beta_{-}+k\beta_{-}}},\qquad
\beta_{-}=\frac{\beta}{1-(k+K) \beta}
\end{equation}
which yields the expected result $a_{-}=-a$. Thus,
the relation between $v_{-}$ and $v$ derived above using
(\ref{In3}) (similar relations are commonly obtained using kinematic arguments
\cite{WinI}) may be considered as resulting from the group
property of the transformations.

For deriving consequences of the transformations it is convenient to write the inverse
transformations in terms of $\beta$ (not $\beta_{-}$), as follows
\begin{eqnarray}\label{TrAnisInvP}
&&X=\frac{R^{-1}}{\sqrt{\left(1-K\beta\right)^2-\beta^2}}
\left(x\left(1-K\beta-k\beta\right)+c t \beta\right),\nonumber \\
&&cT=\frac{R^{-1}}{\sqrt{\left(1-K\beta\right)^2-\beta^2}} \left(c
t+x\left(\left(1-K^2\right)\beta+K-k\right)\right)\\
\label{YZAnis2InvP}&&Y=R^{-1}y,\quad Z=R^{-1}z
\end{eqnarray}

The formulas for the velocity transformation are readily obtained from (\ref{TrAnisP}) and
(\ref{YZAnis2P}), as follows
\begin{eqnarray}\label{VelTr1}
&&u_x=\frac{c(U_X-c \beta)}{Q},\; u_y=\frac{c U_Y
\sqrt{\left(1-K\beta\right)^2-\beta^2}}{Q},\;
u_z=\frac{c U_Z \sqrt{\left(1-K\beta\right)^2-\beta^2}}{Q},\nonumber\\
&& Q=c(1- K \beta)+U_X(K^2\beta-K-\beta)+k(U_X-c \beta)
\end{eqnarray}
where $(U_X,U_Y,U_Z)$ and $(u_x,u_y,u_z)$ are the velocity components in the frames $S$ and $S'$ respectively.
Remark that the relation (\ref{GrProp}) derived above represents the properly specified
first equation of (\ref{VelTr1}).

The transformations (\ref{VarPhi}) --
(\ref{YZAnis2P}) contain an indefinite function $k(a)$. The scale factor $R$
also depends on that function. The transformations
are specified in the next section.

\section{Specifying the transformations
}

In the derivation of the transformations in the previous section, the arguments, that there exists a preferred frame in which the light speed is isotropic and that the anisotropy of the one-way speed of light in a specific frame is due to  its motion relative to the preferred frame,
have not been used.
In the framework of the derivation, nothing distinguishes the frame in which $k=0$ from others and the transformations from/to that frame are members of a group of transformations that are equivalent to others.  Thus, the theory developed above is a counterpart of the standard special relativity kinematics which incorporates an anisotropy of the light propagation, with the anisotropy parameter varying from frame to frame. Below the transformations between inertial frames derived in Section 2 are specified   based on that anisotropy of the one-way speed of light in an inertial  frame is caused by its motion with respect to the preferred frame.

First, this leads to the conclusion that the anisotropy parameter $k_s$ in an arbitrary frame $s$ moving with respect to the preferred frame with velocity $\bar{v_s}$ should be given by some (universal) function $k_s=F\left(\bar{\beta_s}\right)$ of that velocity. Equations (\ref{LEAnis0}) and (\ref{GrPAnisP}) imply that $k=k\left(a\left(\beta,K\right),K\right)$  which being specified for the transformation from the preferred frame to the frame $s$ by setting $K=0,\;k=k_s,\;\beta=\bar{\beta_s}$ yields $k_s=F\left(\bar{\beta_s}\right)$. (It could be expected, in general, that a size of the anisotropy depends on the velocity relative to the preferred frame but, in the present analysis,  it is not a presumption but a part of the framework.)

Next, consider three inertial reference frames $\bar{S}$, $S$ and $S'$. As in the preceding analysis,
the standard configuration, with the $y$- and $z$-axes of the three
frames being parallel and the relative motion being along the
common $x$-axis (and along the direction of the anisotropy vector), is assumed. The space and time coordinates and the anisotropy parameters in the frames $\bar{S}$, $S$ and $S'$ are denoted respectively as $\{\bar{x},\bar{y},\bar{z},\bar{t},\bar{k}\}$, $\{X,Y,Z,T,K\}$ and $\{x,y,z,t,k\}$. The frame $S'$ moves relative to $S$ with velocity $v$ and velocities of the frames $S$ and $S'$ relative to the frame $\bar{S}$ are respectively $\bar{v_1}$ and $\bar{v_2}$. A relation between $\bar{v_2}$, $v$ and $\bar{v_1}$ can be obtained from the equation expressing a group property of the transformations, as follows
\begin{equation}\label{S1}
a_2=a_1+a
\end{equation}
where $a_2$, $a_1$ and $a$ are the values of the group parameter corresponding to the transformations from $\bar{S}$ to $S'$, from $\bar{S}$ to $S$  and from $S$ to $S'$ respectively. Those values are expressed through the velocities and the anisotropy parameter values by a properly specified equation (\ref{GrPAnisP}) which, upon substituting into equation (\ref{S1}), yields
\begin{equation}\label{S2}
\frac{1}{2}\ln{\frac{1+\bar{\beta_2} - \bar{k}\bar{\beta_2}}{1-\bar{\beta_2}-\bar{k}\bar{\beta_2}}}=\frac{1}{2}\ln{\frac{1+\bar{\beta_1} - \bar{k}\bar{\beta_1}}{1-\bar{\beta_1}-\bar{k}\bar{\beta_1}}}+\frac{1}{2}\ln{\frac{1+\beta - K\beta}{1-\beta-K\beta}}
\end{equation}
where
\begin{equation}\label{S3}
\bar{\beta_2}=\frac{\bar{v_2}}{c},\quad \bar{\beta_1}=\frac{\bar{v_1}}{c},\quad \beta=\frac{v}{c}
\end{equation}
Exponentiation of equation (\ref{S2}) yields
\begin{equation}\label{S4}
\bar{\beta_2}=\frac{\bar{\beta_1}+\beta\left(1-\left(\bar{k}+K\right)\bar{\beta_1}\right)}
{1+\beta\left(\bar{k}-K+\left(1-\bar{k}^2\right)\bar{\beta_1}\right)}
\end{equation}

Let us now choose the frame $\bar{S}$ to be a preferred frame.
Then, $\bar{k}=0$ and for the frames $S$ and $S'$ we have
\begin{equation}\label{S5}
K=F\left(\bar{\beta_1}\right),\quad k=F\left(\bar{\beta_2}\right)
\end{equation}
With $\bar{\beta_s}=f\left(k_s\right)$ being a function inverse to $F\left(\bar{\beta_s}\right)$, using in (\ref{S4}) the equalities inverse to those of (\ref{S5}) together with $\bar{k}=0$ yields
\begin{equation}\label{S6}
f\left(k\right)=\frac{f\left(K\right)+\beta\left(1-K f\left(K\right)\right)}
{1+\beta\left(-K+f\left(K\right)\right)}
\end{equation}
If the function $f\left(k_s\right)$ were known, the relation (\ref{S6}), that implicitly defines the anisotropy parameter $k$ in the frame $S'$ as a function of the anisotropy parameter $K$ in the frame $S$ and the relative velocity $v$ of the frames, would provide a formula for the transformation of the anisotropy parameter $k$. This would allow to specify the transformations (\ref{TrAnisP}) and (\ref{YZAnis2P}) by substituting that formula for $k$ into the equation of transformation for $t$ and calculating the scale factor $R$ using that formula with $\beta$ expressed as a function of a group parameter $a$ from (\ref{GrPAnisP}).

Although the function $F\left(\bar{\beta_s}\right)$ is not known, a further specification can be made based on the argument that an expansion of the function $F\left(\bar{\beta_s}\right)$ in a series with respect to $\bar{\beta_s}$ should not contain a quadratic term since it is expected that a direction of the anisotropy vector changes to the opposite if a direction of a motion with respect to a preferred frame is reversed: $F\left(\bar{\beta_s}\right)=-F\left(-\bar{\beta_s}\right)$. Thus, with accuracy up to the third order in $\bar{\beta_s}$, the dependence of the anisotropy parameter on the velocity with respect to a preferred frame can be approximated by
\begin{equation}\label{S7}
k_s=F\left(\bar{\beta_s}\right)\approx q \bar{\beta_s}, \quad \bar{\beta_s}=f\left(k_s\right)\approx k_s/q
\end{equation}
Introducing the last equation of (\ref{S7}) into (\ref{S6}) yields
\begin{equation}\label{S8}
k=\frac{q\left(K+\beta\left(q-K^2\right)\right)}{q+\beta K\left(1-q\right)}
\end{equation}
which is the expression to be substituted for $k$ into (\ref{TrAnisP}).
To calculate the scale factor in (\ref{TrAnisP}) and (\ref{YZAnis2P}), $\beta$ is expressed as a function of a group parameter $a$ from
(\ref{GrPAnisP}), as follows
\begin{equation}\label{S9}
\beta=\frac{\sinh{a}}{K\sinh{a}+\cosh{a}}
\end{equation}
which, being substituted into (\ref{S8}), yields
\begin{equation}\label{S10}
k(a)=\frac{q\left(K \cosh{a}+q \sinh{a}\right)}{K\sinh{a}+q\cosh{a}}
\end{equation}
Then using (\ref{S10}) in (\ref{VarPhi}), with (\ref{GrPAnisP}) substituted for $a$ in the result,  yields
\begin{equation}\label{S11}
R=\left(\frac{q^2\left(1+\beta\left(1-K\right)\right)\left(1-\beta\left(1+K\right)\right)}
{\left(q+\beta K\left(1-q\right)\right)^2}\right)^{\frac{q}{2}}
\end{equation}

Thus, after the specification, the transformations between inertial frames incorporating anisotropy of light propagation are defined by equations (\ref{TrAnisP}) and (\ref{YZAnis2P}) with $k$ given by (\ref{S8}) and the scale factor given by (\ref{S11}). It is readily checked that the specified transformations satisfy the correspondence principle. All the equations contain only one undefined parameter, a universal constant $q$.

It should be clarified that, although the specification relies on the approximate relation (\ref{S7}), the transformations, with $k$ and $R$ defined by (\ref{S8}) and (\ref{S11}), are \textit{not} approximate and they do possess the group property.  The transformations (\ref{TrAnisP}) and (\ref{YZAnis2P}) form a group, even with $k(a)$ (or $k(K,\beta)$) undefined, provided that the transformation of $k$ obeys the group property. Since the relation (\ref{S6}), defining that transformation, is  a particular case of the relation (\ref{S4}) obtained from equation (\ref{S1}) expressing the group property, the transformation of $k$ satisfies the group property with any form of the function $F(\beta_s)$, and, in particular, with that defined by (\ref{S7}). Nevertheless, a straightforward check can be made that the specified transformation (\ref{S10}) obeys the group properties. Using the notation
\begin{equation}\label{S12}
\kappa(a,k)=\frac{q\left(k \cosh{a}+q \sinh{a}\right)}{k\sinh{a}+q\cosh{a}}
\end{equation}
and introducing, in addition to $S$ and $S'$, the frame $S_0$ with the anisotropy parameter $k_0$, one can check that
\begin{equation}\label{S13}
\kappa\left(a,\kappa\left(a_0,k_0\right)\right)=\kappa\left(a+a_0,k_0\right)
\end{equation}
Similarly it is readily verified that $\kappa\left(-a,\kappa \left(a,k\right)\right)=k$ and $\kappa(0,k)=k$.
Alternatively, one can calculate the group generator $\chi(k)$ as
\begin{equation}\label{S14}
\chi(k)=\frac{\partial \kappa (a,k)}{\partial a}\biggl\vert_{a=0}=q-\frac{k^2}{q}
\end{equation}
and solve the initial value problem
\begin{equation}\label{S15}
\frac{d k(a)}{d
a}=q-\frac{k(a)^2}{q},\quad k(0)=K
\end{equation}
to be assured that it, as expected, yields (\ref{S10}). Thus, as a matter of fact, what is specified using the approximate relation  (\ref{S7}) is the form of the group generator $\chi(k)$ in the group of transformations defined on the basis of the first principles.

\section{Consequences of the transformations}

\noindent \textit{Length contraction}. Consider a rod that is at rest along the $x$-axis in the frame
$S'$ with the coordinates of its ends being $x_1$ and $x_2$. In
order to obtain its length in the frame $S$ one has to measure the
coordinates of its front tip $X_1$ and of its end $X_2$ at the
same time moment $T_1=T_2$. Using the transformations
(\ref{TrAnisP}) we have
\begin{equation}\label{Length1V}
x_1-x_2=\frac{R}{\sqrt{\left(1-K\beta\right)^2-\beta^2}}(X_1-X_2)
\end{equation}
So we obtain the length contraction relation in the form
\begin{equation}\label{Length1aV}
L=L'\left(R^{-1}
\sqrt{\left(1-K\beta\right)^2-\beta^2}\right)
\end{equation}
Note that, in the presence of the
anisotropy, the terms "length contraction" and "time dilation"
become conditional in a sense. In general, it could be, for
example, length dilation rather than length-contraction but, as it
is commonly accepted in the literature, the corresponding
relations are referred to as the length-contraction and
time-dilation relations.


\smallskip

\noindent \textit{Time dilation}.
Consider a clock $C'$ placed at
rest in $S'$ at a point on the $x$-axis with the coordinate
$x=x_1$. When the clock records the times $t=t_1$ and $t=t_2$ the
clock in $S$ which the clock $C'$ is passing by at those moments
will record times $T_1$ and $T_2$ given by the transformations
(\ref{TrAnisInvP}) where it should be evidently set $x_2=x_1$.
Subtracting the two relations we obtain the time dilation relation
\begin{equation}\label{Time1V}
\Delta
T=\frac{R^{-1}}
{\sqrt{\left(1-K\beta\right)^2-\beta^2}}\Delta
t
\end{equation}
If clock were at rest in the frame $S$ the time dilation relation would be
\begin{equation}\label{Time1Va}
\Delta
t=\frac{R\left(1-K \beta-k \beta\right)}
{\sqrt{\left(1-K\beta\right)^2-\beta^2}}\Delta
T=\frac{R}
{\sqrt{\left(1-k\beta_{-}\right)^2-\beta_{-}^2}}\Delta
T
\end{equation}
with $\beta_{-}$ defined by (\ref{VelRec1Kc}).

\smallskip

\noindent \textit{Aberration law}.
The light aberration law can be derived using the formulas (\ref{VelTr1}) for the velocity transformation.
The relation between directions of a light ray in the two inertial frames
$S$ and $S'$ is obtained by setting $U_X=c \cos\Theta/(1+K \cos\Theta)$ and $u_x=c \cos\theta/(1+k
\cos\theta)$ in the first equation of (\ref{VelTr1}). Then solving for $\cos\theta$ yields
\begin{equation}\label{Aber1} \cos\theta=\frac{\cos\Theta-\beta
(1+K\cos\Theta)}{1-\beta (\cos\Theta+K)}
\end{equation}
where $\theta$ and $\Theta$ are the angles between the direction
of motion and that of the light propagation in the frames of a
moving observer (the Earth) and of an immovable source (star or galaxy)
respectively.
(Equation (\ref{Aber1}) could be obtained in several other ways, for example,
straight from the transformations (\ref{TrAnisP}) and
(\ref{YZAnis2P}) by rewriting them in spherical coordinates and
then specifying to radial light rays.) Introducing
$\tilde\theta=\theta -\pi$ and $\tilde\Theta=\Theta -\pi$ as the
angles between the direction of motion and the line of sight one
gets the aberration law
\begin{equation}\label{Aber}
\cos\tilde\theta=\frac{\cos\tilde\Theta+\beta
(1-K\cos\tilde\Theta)}{1+\beta (\cos\tilde\Theta-K)}
\end{equation}

\smallskip

\noindent \textit{Doppler effect}.
Consider a source of electromagnetic radiation (light) in
a reference frame $S$ very far from the observer
in the frame $S'$ moving with
velocity $v$ with respect to $S$ along the $X$-axis with $\Theta$ being the angle between the direction of the observer
motion and that of the light propagation as measured in a frame of
the source. Let two pulses of the radiation are emitted from the source with the time interval $(\delta T)_e$ (period). Then the interval $(\delta T)_r$ between the times of arrival of the two pulses to the observer, as measured by a clock in the frame of the source $S$, is
\begin{equation}\label{DE1}
(\delta T)_r=(\delta T)_e+\frac {\delta L}{V}
\end{equation}
where $\delta L$ is a difference of the distances traveled by the two pulses, measured in the frame of the source $S$, and $V$ is the speed of light in the frame $S$ given by
\begin{equation}\label{DE2}
\delta L=v (\delta T)_r\cos{\Theta}, \quad V=\frac{c}{1+K \cos{\Theta}}
\end{equation}
Substituting (\ref{DE2}) into (\ref{DE1}) yields
\begin{equation}\label{DE3}
(\delta T)_e=(\delta T)_r\left(1-\beta \cos{\Theta}\left(1+K \cos{\Theta}\right)\right)
\end{equation}
The interval $(\delta t)_r$ between the moments of receiving the two pulses by the observer in the frame $S'$, as measured by a clock at rest in $S'$, is related to $(\delta T)_r$ by the time dilation relation (\ref{Time1V}), as follows
\begin{equation}\label{DE4}
(\delta T)_r=\frac{R^{-1}}
{\sqrt{\left(1-K\beta\right)^2-\beta^2}}(\delta t)_r
\end{equation}
Thus, the periods of the electromagnetic wave measured in the frames of the source and the receiver are related by
\begin{equation}\label{DE5}
(\delta T)_e=\frac{R^{-1}\left(1-\beta \cos{\Theta}\left(1+K \cos{\Theta}\right)\right)}
{\sqrt{\left(1-K\beta\right)^2-\beta^2}}(\delta t)_r
\end{equation}
so that the relation for the frequencies is
\begin{equation}\label{DE6}
\nu_r=\nu_e \frac{R^{-1}\left(1-\beta \cos{\Theta}\left(1+K \cos{\Theta}\right)\right)}
{\sqrt{\left(1-K\beta\right)^2-\beta^2}}
\end{equation}
where $\nu_e$ is the emitted wave frequency and $\nu_r$ is the wave frequency measured by the observer moving with respect to the source. (This formula could be derived in several other ways, for example, using the condition of  invariance of the wave phase.)

To complete the derivation of the formula for the Doppler shift, the relation (\ref{DE6}) is to be transformed such that the angle $\theta$ between the wave vector and the direction of motion measured in the frame of the observer $S'$ figured instead of $\Theta$ which is the corresponding angle measured in the frame of
the source. Using the aberration formula (\ref{Aber1}), solved for $\cos{\Theta}$, as follows
 \begin{equation}\label{DE7} \cos\Theta=\frac{\cos\theta+\beta
(1-K\cos\theta)}{1+\beta (\cos\theta-K)}
\end{equation}
in the relation (\ref{DE6}) yields
\begin{equation}\label{DE9}
\nu_r=\nu_e \frac{R^{-1}\left(1+\beta \cos{\theta}\left(1-K \cos{\theta}\right)\right)\sqrt{\left(1-K\beta\right)^2-\beta^2}}
{\left(1-K \beta +\beta \cos{\theta}\right)^2}
\end{equation}
Finally, introducing the angle $\tilde\theta=\theta -\pi$ between the line of sight and the direction of the observer motion one obtains the relation for a shift of frequencies due to the Doppler effect in the form
\begin{equation}\label{DE10}
\nu_r=\nu_e \frac{R^{-1}\left(1-\beta \cos{\tilde\theta}\left(1+K \cos{\tilde\theta}\right)\right)\sqrt{\left(1-K\beta\right)^2-\beta^2}}
{\left(1-K \beta -\beta \cos{\tilde\theta}\right)^2}
\end{equation}

\section{Cosmological implications}

According to the modern view, there exists a preferred frame of reference related to the cosmic microwave background (CMB), more precisely to the last scattering surface (LSS). Let us apply the equations of  the anisotropic special relativity developed above to describe effects caused by an observer motion (our galaxy's peculiar motion) with respect to the CMB. Using equations of the standard special relativity in that context is inconsistent. The standard relativity theory framework
is in contradiction with existence of a preferred frame while the anisotropic special relativity developed in the present paper naturally combines the special relativity principles with the existence of a preferred frame.
 In order to apply equations of the anisotropic special relativity for describing the physical phenomena  in a frame moving with respect to the LSS, let choose
 the frame $S$ to be a preferred frame and the frame $S'$ to be a frame of an observer moving with respect to the preferred frame.
 Then the coordinate transformations from the preferred frame $S$ to the  frame $S'$ of the moving observer are obtained by setting $K=0$ in equations (\ref{TrAnisP}), (\ref{YZAnis2P}), (\ref{S11}) and (\ref{S8}) which yields
\begin{eqnarray}\label{PF1}
&&x=\left(X-c T\beta\right)\left(1-\beta^2\right)^{\frac{q-1}{2}},
\quad c t=\left(cT\left(1-q\beta^2\right)
-Z\beta\left(1-q\right)\right)\left(1-\beta^2\right)^{\frac{q-1}{2}}\nonumber\\
&&y=Y\left(1-\beta^2\right)^{\frac{q}{2}},\quad z=Z\left(1-\beta^2\right)^{\frac{q}{2}}
\end{eqnarray}
where $q$ is a universal constant.
Equation of aberration of light (\ref{Aber}) with $K=0$
converts into the common aberration law of the standard theory
\begin{equation}\label{PF2}
\cos\tilde\theta=\frac{\cos\tilde\Theta+\beta
}{1+\beta \cos\tilde\Theta}
\end{equation}
while equation (\ref{DE6}), describing the Doppler frequency shift for the light emitted at the LSS
and received by an observer moving with respect to the LSS, differs from its counterpart of the standard relativity by the factor $R^{-1}$, as follows
\begin{equation}\label{PF3}
\nu_r=\nu_e \frac{R^{-1}\left(1-\beta \cos{\Theta}\right)}
{\sqrt{1-\beta^2}}
\end{equation}
The inverse $R^{-1}$ of (\ref{S11}) for $K=0$ takes the form
\begin{equation}\label{PF3a}
R^{-1}=\left(1-\beta^2\right)^{-\frac{q}{2}}
\end{equation}
Substituting (\ref{PF3a}) into (\ref{PF3}) yields
\begin{equation}\label{PF3b}
\nu_r=\nu_e \left(1-\beta^2\right)^{-\frac{1}{2}-\frac{q}{2}}\left(1-\beta \cos{\Theta}\right)
\end{equation}
 Thus,  in terms of the angle $\Theta$ between the direction of the observer
motion and that of the light propagation as measured in a frame of
the source, the Doppler frequency shift is a pure dipole pattern as it is in the standard relativity. However, the amplitude of the shift includes an additional factor which depends on the value of the universal constant $q$.

Equation (\ref{DE10}) incorporating the effect of light aberration and thus relating the frequency $\nu_e$ of the light emitted at the LSS to the frequency $\nu_r$ measured by an observer moving with respect to the LSS, with the use of (\ref{PF3a}) becomes
\begin{equation}\label{PF4}
\nu_r=\nu_e \frac{\left(1-\beta^2\right)^{\frac{1}{2}-\frac{q}{2}}}{1-\beta \cos{\tilde \theta}}
\end{equation}
where  $\tilde\theta$ is the angle between the line of sight and the direction of the observer motion as measured in the frame of the observer. In the context of the CMB anisotropy, one should switch from the frequencies to effective thermodynamic temperatures of the CMB blackbody radiation using the relation \cite{PW}
\begin{equation}\label{PF5}
\frac{T(\tilde\theta)}{\nu_r}=\frac{T_0}{\nu_e}
\end{equation}
where $T_0$ is the effective temperature measured by the observer, that is at rest relative to the LSS and sees strictly isotropic blackbody radiation, and $T(\tilde\theta)$ is the effective temperature of the blackbody radiation for the moving observer looking in the fixed direction $\tilde\theta$.
Substituting (\ref{PF4}) into (\ref{PF5}) yields
\begin{equation}\label{PF6}
T(\tilde\theta)=T_0 \frac{\left(1-\beta^2\right)^{\frac{1}{2}-\frac{q}{2}}}{1-\beta \cos{\tilde \theta}}
\end{equation}
Thus, the angular distribution of the CMB effective temperature seen by an observer moving with respect to the CMB is not altered by the light speed anisotropy. However, the anisotropy influences the mean temperature which now does not coincide with the temperature $T_0$ measured by the observer, that is at rest relative to the LSS, but differs from it by the factor $\left(1-\beta^2\right)^{-\frac{q}{2}}$.  Developing equation (\ref{PF6}) up to the second order in $\beta$ yields
\begin{equation}\label{PF7}
T(\tilde\theta)=T_0 \left(1+q\frac{\beta^2}{2}+\beta \cos{\tilde \theta}+\frac{\beta^2}{2}\cos{2\tilde \theta}\right)
\end{equation}
which implies that, up to the order $\beta^2$, the amplitudes of the dipole and quadrupole patterns remain the same, only the constant term is modified.

It is worth reminding that, even though the specified law (\ref{S7})  is linear in $\beta$, it does include the second order term which is identically zero. Thus, describing the anisotropy effects, which are of the order of $\beta^2$, by equations (\ref{PF6}) and (\ref{PF7}) is legitimate.

Equations (\ref{DE10}) and  (\ref{S11}) can be used to derive the Doppler frequency shift in the case when an observer in the frame moving with respect to the CMB (Earth) receives light from an object (galaxy) which is also moving with respect to that preferred frame. The present formulation, which assumes that all motions are along the same axis, implies that the relative motion of the object and the observer is along the direction of the observer motion relative to the CMB. It is straightforward, within the framework developed in the present study,  to extend the analysis to defining a group of transformations which is not restricted by that assumption.
It is worthwhile to note that the results related to the CMB temperature distribution considered above are not influenced by that assumption since only two frames, moving and preferred frames, figure in the analysis.

To obtain a formula convenient for applications some alterations are needed. First, the frequency shift is to be expressed in terms of the velocity $v_g$  (or $\beta_g$) of the object relative to the observer which, in the presence of anisotropy, is not equal to the velocity $v$ (or $\beta$) of the observer with respect to the object that figures in equations (\ref{DE10}) and  (\ref{S11}). To do this  $\beta$ is expressed through $\beta_g$ from equation (\ref{VelRec1Kc}) where $\beta_{-}$ stays for $\beta_g$, as follows
\begin{equation}\label{PF8}
\beta=\frac{\beta_g}{1+\left(K+k\right)\beta_g}
\end{equation}
Next, it is needed to express the frequency shift in terms of the observer velocity relative to the CMB $\bar{v}$ but not of the object velocity relative to the CMB. Since the velocities with respect to the CMB are related to the anisotropy parameters by (\ref{S7}) it implies that the anisotropy parameter in the frame of the observer $K$ is to be expressed through $k$. It is done using equation (\ref{S8}) with $\beta$ replaced by (\ref{PF8}) which yields
\begin{equation}\label{PF9}
K=\frac{q\left(k+\left(k^2-q\right)\beta_g\right)}{q+\left(q-1\right)k\beta_g}
\end{equation}
The final formula for the frequency shift is obtained from equations (\ref{DE10}) and  (\ref{S11}) by substituting (\ref{PF8}) for $\beta$ and next substituting (\ref{PF9}) for $K$ and expressing $k$ as $k=q \bar{\beta}$ afterwards. In order not to complicate matters, the case when the object motion is along the line of sight, $\tilde{\theta}=0$ is considered and the resulting expression for the frequency shift is expanded up to the order of $\beta_g^2$ and $\bar{\beta}\beta_g$ which yields
\begin{equation}\label{PF10}
\nu_r=\nu_e\left(1+\left(1+q \bar{\beta}\right)\beta_g+\frac{1}{2}\left(1-q\right)\beta_g^2\right)
\end{equation}
Thus, corrections to the Doppler shift due to the presence of the anisotropy (the terms multiplied by $q$ in (\ref{PF10})) are of the second order in velocities.

\section{Concluding comments}

A counterpart of the special relativity kinematics has been developed to remedy the situation
when the principles of special relativity
are in contradiction with a commonly accepted view that there exists  a preferred universal rest frame, that of the cosmic background radiation. Analysis of the present paper shows that, despite the general consensus that the special relativity principles should be abolished if the existence of the preferred frame is accepted, a synthesis of those seemingly incompatible concepts is possible. 
The framework developed
doesn't abolish the basic principles of special relativity but simply uses the freedom in applying those principles in order to incorporate a preferred frame into the theory. A degree of anisotropy of the one-way velocity, which  is commonly considered as irreducibly conventional, acquires meaning of a characteristic of the really existing anisotropy caused by motion of an inertial frame relative to the preferred frame. In that context, the fact, that there exists the inescapable entanglement between remote clock synchronization and one-way speed of light, does not imply conventionality of the one-way velocity but means  that the synchronization procedure is to be made using the one-way velocity selected by the size of the really existing anisotropy (like as the Einstein synchronization using the isotropic one-way velocity is selected in the case of an isotropic system).


Incorporating the anisotropy of the one-way speed of light into the framework based on the relativity principle and the principle of constancy of the two-way speed of light yields equations differing from those of the standard relativity. The deviations depend on the value of an universal constant $q$ where $q=0$ corresponds to the standard relativity theory with the isotropic one-way speed of light and Einstein synchronization in all the frames. The measurable effects following from the theory equations can be used to validate the theory and provide estimates for $q$.
From somewhat different perspective, it means that, even though direct measuring of the one-way speed of light is not possible,
the anisotropy of the one-way speed of light may reveal itself in measurable effects.

Applying the theory to
the problem of calculating
the CMB temperature distribution is conceptually attractive since it removes the inconsistency of the usual approach when formulas of the standard special relativity, in which a preferred frame is not allowed, are applied to define effects caused by motion with respect to the preferred frame. 
Nevertheless, other measurable effects,
in particular, the Doppler frequency shift measured by the Earth  observer receiving light from an object that is also moving relative to the GMB frame, might be easier to identify. The object could also be a  light emitter in laboratory experiments.

The constant $q$ has a definite physical meaning of the coefficient in the formula (\ref{S7}) defining dependence of the parameter $k$ of anisotropy of the one-way speed of light in a particular frame on the frame velocity with respect to a preferred frame. Nevertheless a direct measuring of that constant is not possible since no experiment is a "one-way  experiment". (We leave aside a discussion of the papers in which measuring of the one-way
speed of light is reported, as well as of the papers refuting them.) The present theory provides the possibility of obtaining estimates for that fundamentally important constant.
It is worthwhile to note that even though it were found that the constant $q$ is very small, which would mean that applying the present theory yields results practically identical to those of the standard relativity, this would not reduce the importance of the present framework which reconciles the principles of special relativity with the existence of the privileged CMB frame. As a matter of fact, it would justify the application of the standard relativity in that situation.


\appendix

\section{Equation of light propagation}

We will define the form of the equation of light propagation based
on the law (\ref{In4}) for the light speed variation. If we use
the spherical coordinate system
\begin{equation}\label{Spher0}
x=r \cos \theta,\quad y=r \sin \theta \sin \phi,\quad z=r \sin
\theta \cos \phi
\end{equation}
with the axis $x$ directed along the anisotropy vector
$\mathbf{k}$, then the angle $\theta_k$ in (\ref{In4}) coincides
with the polar angle $\theta$ so that the law of the variation of
speed of light in space becomes
\begin{equation}\label{VLAnis}
V=\frac{c}{1+k\cos \theta}
\end{equation}

To derive the equation for light propagation corresponding to the
law (\ref{VLAnis}) we start from
\begin{equation}\label{metr0}
g_{ik}dx^{i}dx^{k}=0;
\end{equation}
with $i$ and $k$ running from $0$ to $3$ ($g_{00}>0$) and $x^0=c
t$, $x^1=x$, $x^2=y$, $x^3=z$. To define $g_{ik}$ such that
(\ref{metr0}) corresponded to the law (\ref{VLAnis}) we will use
the expression for the light velocity (see, e.g., \cite{Moller1}):
\begin{equation}\label{LightVel1}
V^{\alpha}=\frac{dx^{\alpha}}{dt}=V n^{\alpha};\quad
V(n^{\alpha})=\frac{c \sqrt{g_{00}}}{1+\gamma_{\mu}n^{\mu}};\qquad
\gamma_{\mu}=-\frac {g_{\mu 0}}{\sqrt{g_{00}}}
\end{equation}
where Greek letters run from $1$ to $3$ as distinct from Latin
letters that run from 0 to 3.. We will also use the relation
\begin{equation}\label{LightVel2}
\gamma_{\mu \nu}n^{\mu}n^{\nu}=1;\quad \gamma_{\mu \nu}=-g_{\mu
\nu}+\gamma_{\mu}\gamma_{\nu}
\end{equation}
Based on the symmetry of the problem we have
\begin{equation}\label{LightVel3}
g_{20}=g_{30}=0,\; g_{22}=g_{33}=-1\quad \Rightarrow
\gamma_{2}=\gamma_{3}=0,\; \gamma_{22}=\gamma_{33}=1
\end{equation}
Then it follows from (\ref{LightVel1}), (\ref{VLAnis}) and
(\ref{LightVel2}) that
\begin{equation}\label{LightVel4}
g_{00}=1,\quad -g_{10}n^{1}=k \cos \theta;\qquad
(-g_{11}+g_{10}^2)(n^{1})^2+(n^{2})^2+(n^{3})^2=1
\end{equation}
With $(n^{1}=\cos \theta,\; n^{2}=\sin \theta \sin \phi,\;
n^{3}=\sin \theta \cos \phi)$ we obtain $g_{10}=-k$ and
$g_{11}=k^2-1$ so that the equation for light propagation becomes
\begin{equation}\label{LPAnis}
c^2 dt^2-2k c\; dtdx-(1-k^2)dx^2-dy^2-dz^2=0
\end{equation}

Thus, although equation (\ref{LPAnis}) (with $dy=dz=0$) commonly
arises in the traditional one-dimensional arguments it corresponds
to the three-dimensional law (\ref{VLAnis}). It can be also
demonstrated by rewriting (\ref{LPAnis}) in the spherical
coordinates (\ref{Spher0}) for light rays propagating in radial
direction, as follows
\begin{equation}\label{LPAnisRad}
c^2 dt^2-2k c\cos \theta\; dtdr-(1+k \cos\theta)(1-k
\cos\theta)dr^2=0
\end{equation}
Solving (\ref{LPAnisRad}) for $V=\frac{dr}{dt}$ yields two roots
\begin{equation}\label{LPAnisRad1}
V_{+}=\frac{c}{1+k\cos\theta},\quad V_{-}=-\frac{c}{1-k\cos\theta}
\end{equation}
corresponding to two different directions of the light propagation
according to the law (\ref{VLAnis}).

\section{Derivation of the Lorentz transformations based on invariance of the equation of light
propagation, group property and correspondence principle}

As in Section 2, two arbitrary inertial reference frames $S$ and $S'$ in
the standard configuration, with the $y$- and $z$-axes of the two
frames being parallel while the relative motion is along the
common $x$-axis, are considered. The space and time coordinates in $S$ and $S'$
are denoted respectively as $\{X,Y,Z,T\}$ and $\{x,y,z,t\}$. The
velocity of the $S'$ frame with respect to
$S$ is denoted by $v$. Transformations between the frames are
derived based on the same first principles as in Section 2: Invariance of the equation of light
propagation; Group property; Correspondence principle.

Equations of light propagation in the frames $S$ and $S'$ are
\begin{eqnarray}
&&\label{2.b1A}
 c^2 dT^2-dX^2-dY^2-dZ^2=0,\\
&&\label{2.b1B} c^2 dt^2-dx^2-dy^2-dz^2=0
\end{eqnarray}
The coordinate and time
transformations between inertial frames are sought from the conditions that they leave the equation of light propagation invariant (transform equation (\ref{2.b1A}) into (\ref{2.b1B})) and form a one-parameter group with the group parameter $a=a(v)$ (such that $v\ll 1$ corresponds
to $a\ll 1$):
\begin{equation}\label{2.4b}
x=f(X,T;a), \; t=q(X,T;a),\quad y=g(Y,Z;a), \; z=h(Y,Z;a)
\end{equation}

It is worth remarking that the linearity assumption is not
imposed.

The Lie infinitesimal technique is used, as in Section 2.
The infinitesimal transformations corresponding to (\ref{2.4b}) are
\begin{equation}\label{2.5b}
x\approx X+\xi(X,T)a, \quad t\approx T+\tau(X,T)a,\quad y\approx
Y+\eta(Y,Z)a, \quad z\approx Z+\zeta(Y,Z)a
\end{equation}
Substituting the infinitesimal transformations
(\ref{2.5b}) into equation (\ref{2.b1B}) with
subsequent linearizing with respect to $a$ and using
equation (\ref{2.b1A}) to eliminate $dT^2$ yields
\begin{equation}\label{DE0b}
(\tau_T-\xi_X)dX^2+(\tau_T-\eta_Y)dY^2+(\tau_T-\zeta_Z)dZ^2
+(c^2\tau_X-\xi_T)dX dT-(\eta_Z+\zeta_Y)dY dZ=0
\end{equation}
The correspondence principle is applied to specify
the group generator $\xi(X,T)$ to the form (\ref{2.6}).
Then solving the determining equations for the generators $\tau$,
$\eta$ and $\zeta$ yields
\begin{equation}\label{Genb}
\tau=-\frac{b}{c^2}X+c_1;,\quad \eta=\omega Z+c_2,\;\zeta=-\omega
Y+c_3
\end{equation}
where $\omega$, $c_1$, $c_2$ and $c_3$ are arbitrary constants.

Having the infinitesimal group generators defined by (\ref{Genb})
the finite group transformations can be found via solving the Lie
equations with proper boundary conditions. As in Section 2, the
common kinematic restrictions can be imposed to make the constants $c_1$, $c_2$,
$c_3$ and $\omega$ vanishing which eliminates space and time shifts and excludes
rotations in the plane $(y,z)$. The constant $b$ can be eliminated
by redefining the group parameter as $\hat a=a b/c$ (hats will be
omitted in what follows). As the result, the Lie equations take
the forms
\begin{eqnarray}\label{2.7}
\frac{d x(a)}{d a}=-c t(a),\quad \frac{d \left(c
t\left(a\right)\right)}{d
a}=-x(a); \qquad x(0)=X,\quad t(0)=T\\
\label{2.7c}\frac{d y(a)}{d a}=0,\quad \frac{d z(a)}{d a}=0;
\qquad y(0)=Y,\quad z(0)=Z
\end{eqnarray}
The initial data problems (\ref{2.7}) and (\ref{2.7c}) are readily
solved to give
\begin{equation}\label{2.8}
x=X \cosh{a}-cT \sinh{a},\quad c t=c T \cosh{a}-X \sinh{a};\qquad
y=Y,\quad z=Z
\end{equation}
The group
parameter $a$ is related to the velocity $v$ using the
condition
\begin{equation}\label{2.8a}
x=0 \quad \mathrm{for} \quad X=v T
\end{equation}
which yields
\begin{equation}\label{2.8b}
a=\tanh^{-1}\frac{v}{c}\qquad \mathrm{or}\qquad
a=\frac{1}{2}\ln{\frac{1+v/c}{1-v/c}}
\end{equation}
Substitution of (\ref{2.8b}) into (\ref{2.8}) results in the
Lorentz transformations
\begin{equation}
\label{1.1} x=\frac{X-(v/c)c T}{\sqrt{1-v^2/c^2}},\quad c
t=\frac{c T-(v/c) X}{\sqrt{1-v^2/c^2}}\qquad y=Y,\quad z=Z
\end{equation}


\end{document}